\documentclass[prc,twocolumn,showpacs]{revtex4-1}

\usepackage{enumerate}

\usepackage{booktabs}
\usepackage{color}
\usepackage{graphicx}

\usepackage{amsmath}
\usepackage{amssymb}

\def\bvec#1{\mbox{\boldmath $#1$}}

\newcommand{\what}[1]{\widehat{#1}}

\newcommand{\del}[2]{\frac{\partial #1}{\partial #2}}

\newcommand{\bra}{\langle}
\newcommand{\ket}{\rangle}

\newcommand{\idot}{\!\cdot\!}

\newcommand{\beq}{\begin{equation}}
\newcommand{\eeq}{\end{equation}}
\newcommand{\bea}{\begin{eqnarray}}
\newcommand{\eea}{\end{eqnarray}}

\def\fun#1#2{\lower3.6pt\vbox{\baselineskip0pt\lineskip.9pt
\ialign{$\mathsurround=0pt#1\hfil##\hfil$\crcr#2\crcr\sim\crcr}}}

\begin{document}

\title{
  Jost function based on the Hartree-Fock-Bogoliubov formalism
}

\author{K. Mizuyama$^{1}$, N. Nhu Le$^{2,3}$, T. Dieu Thuy$^{2}$, T. V. Nhan Hao$^{2,3}$}

\affiliation{
\textsuperscript{1} Institute of Research and Development, Duy Tan University, Da Nang City, Vietnam \\
\textsuperscript{2} Faculty of Physics, University of Education, Hue University, 34 Le Loi Street, Hue City, Vietnam \\
\textsuperscript{3} Center for Theoretical and Computational Physics, College of Education, Hue University, 34 Le Loi Street, Hue City, Vietnam\\
}


\email{corresponding author: tvnhao@hueuni.edu.vn}
\date{\today}

\begin{abstract}
  We formulate the Jost function formalism based on the Hartree-Fock-Bogoliubov (HFB)
  theory which has been used to represent the nature of the superfluidity of nucleus.
  The Jost function based on the HFB can give the analytic representation of the
  S-matrix for the nucleon elastic scattering targeting on the open-shell nucleus
  taking into account the pairing effect. By adopting the Woods-Saxon potential, we
  show the numerical results of S-matrix poles and their trajectories with
  varying the pairing strength in two cases of Fermi energies: $\lambda=-8.0$ and
  $-1.0$ MeV. The total cross sections in each cases for neutron elastic scattering
  are also analyzed, and we confirmed some staggering shapes or sharp
  resonances originated from the effect of pairing can be seen in the cross section.
\end{abstract}

\maketitle

\section{Introduction}
The coefficient functions of the regular solution of the Schr\"{o}dinger equation
for scattering introduced by Jost~\cite{jost} has been known as the {\it Jost function}.
They have been used to identify the S-matrix poles for the bound states and resonances
of the system on the complex energy/momentum plane.

The S-matrix poles represent the bound states and the resonance energy and width.
The analysis of S-matrix poles has been performed on the scattering data to investigate
the properties of the potential of the target.
An interesting study is to investigate the trajectories of the S-matrix poles on the
complex energy/momentum plane shown as a function of the potential depth~\cite{Hussein}.
It is also possible to analyze the channel coupling effect in terms of the S-matrix poles
on the Riemann sheets defined for each channels~\cite{Rod,Hans}.

The optical potential is indispensable to a reliable extraction of nuclear structure
information from experimental data of various direct nuclear reactions such as
inelastic scattering and transfer reactions. The global optical potentials(GOP) have been
investigated for long time~\cite{watson,Bec,Nada,Var,Hama,Cooper,Koning,Kunieda,Han1,Perey,Dae,Boj,An,Han2},
and succeeded to reproduce the experimental cross section with high accuracy.
The GOP is given by the complex function, the imaginary part gives the absorption of the target
nucleus for the current of the incident beam. The existence of the imaginary part of the GOP has
been justified qualitatively by the Feshbach projection theory\cite{feshbach}.

Recently, the self-consistent particle-vibration coupling (PVC) method has been applied to the
nucleon-nucleus ($NA$) scattering~\cite{mizuyama,mizuyama2,branchon,hao}.
The experimental data for neutron elastic scattering on $^{16}$O~\cite{mizuyama},
proton inelastic scattering on $^{24}$O~\cite{mizuyama2}, neutron elastic scattering on
$^{40}$Ca, its analyzing power~\cite{branchon}, and neutron elastic scattering
on $^{16}$O and $^{208}$Pb~\cite{mizuyama} were successfully reproduced.
In the PVC framework, the microscopic optical potential is given as
the non-local complex potential which gives the proper absorption for the reproduction of the
experimental cross section. These results are consistent with the Feshbach projection theory,
and give the quantitative justification for the existence of the imaginary part of the GOP.

It is well known that there are many sharp peaks are observed at low-energy in the total
and elastic cross section of the $NA$-scattering. Such resonances have been analyzed by
using the R-matrix theory~\cite{rmatrix}, The R-matrix theory is a phenomenological framework
in which reactions of neutrons, charged particles + nuclei are described quantum mechanically
based on boundary conditions for various channels and the scattering matrix (S-matrix) is acquired
from measurement data of cross section. Obtained resonance parameters have been published as
the nuclear data~\cite{jendl,jeff,endf} and used for many applications (Nuclear engineering,
radiotherapy, etc.)
because nuclear resonances are very useful characteristic for the applications.

It was also shown that the sharp resonances appeared in the neutron elastic scattering
can be partially reproduced due to the coupling between the induced neutron and the
giant resonance of the target nucleus~\cite{mizuyama}. This is also consistent with
the Feshbach theory which justified the R-matrix theory and its resonance formula.
Very recently, the multichannel algebraic scattering method (MCAS) which is based on
the phenomenological vibrational model has been applied to study the bound and
resonance properties of the $^{17}$O and $^{17}$F nuclei below and above the nucleon
-core threshold. The narrow resonances of the experimental data could be reproduced
successfully~\cite{amos}.


The results of these recent progress and attempts at low energy $NA$ scattering may
indicate the possibility of a new phenomenological model based on the microscopic
theory of nuclear structure and reaction. However, the pairing correlation has not
been discussed in the direction of these recent progress.

The importance of the pairing correlation has been discussed mainly in the nuclear
structure. The pairing correlation in nuclei is very important to explain the fundamental
properties of nuclei, such as the magic number, the separation energy and the
quadrupole 1st excited state of nucleus, and so on. In the last decades, the investigation of
the pairing effect on nuclear structure has been held based on the microscopic theoretical
framework, such as the Hartree-Fock-Bardeen-Cooper-Schrieffer (HF+BCS)~\cite{hfbcs2017},
the Hartree-Fock-Bogoliubov (HFB)~\cite{hfb2010}, the Highly Truncated
Diagonalization Approach (HTDA)~\cite{htda2002,htda2012} and the Quasiparticle Random Phase
Approximation (QRPA)~\cite{hfb1,qrpa1,qrpaC,qrpaT}. Di-neutron correlation resulting from the
coherent overlap of the continuum states due to the pair correlation at both ground state and
the low-lying excitation of the neutron drip line nuclei has been discussed~\cite{qrpa1}.
Also the quasiparticle resonances of the neutron drip-line nuclei have been investigated
within the framework of HFB\cite{hamamoto, hamamoto2,zhang,kobayashi,grasso,fayans2,michel,Pei,zhang2,oba,zhang3,Sand,betan,sand2}.
The experimental cross section of d($^{9}$Li,$^{10}$Li)p was also analyzed from the point of the
pair resonance based on the HFB formalism~\cite{orrigo2}.

Nevertheless, the role of the pairing correlation for the reaction is not clear yet.
The pairing correlation may have the important role for the reaction and the channel coupling
in which two neutron transfer channel is relevant~\cite{ptrf}.
In order clarify the role of the pairing in the reaction, it may be necessary to
investigate the pairing effect in terms of the S-matrix poles.
The aim of this work is to derive the Jost function formalism based on the HFB formalism
in order to discuss the pairing effect on $NA$-scattering in terms of the
S-matrix poles.

This paper is organized as follows. We introduce the definition of the complex quasiparticle
energy and momentum basing on the HFB equation in Sec.\ref{sec2}.
We show the derivation of the Jost function based on the HFB equation, the definition of the
S-matrix and its unitarity in Sec.\ref{sec3}. In Sec.\ref{sec-num}, we demonstrate the numerical
results adopting the Woods-Saxon potential. The poles of the S-matrix and their trajectories with
varying the paring strength are shown. The appearance of the quasiparticle resonances on the
total cross section of the neutron elastic scattering cross section is also discussed.

\section{Definitions of complex energy \& momentum planes}
\label{sec2}

\begin{figure}[htbp]
\includegraphics[scale=0.25,angle=0]{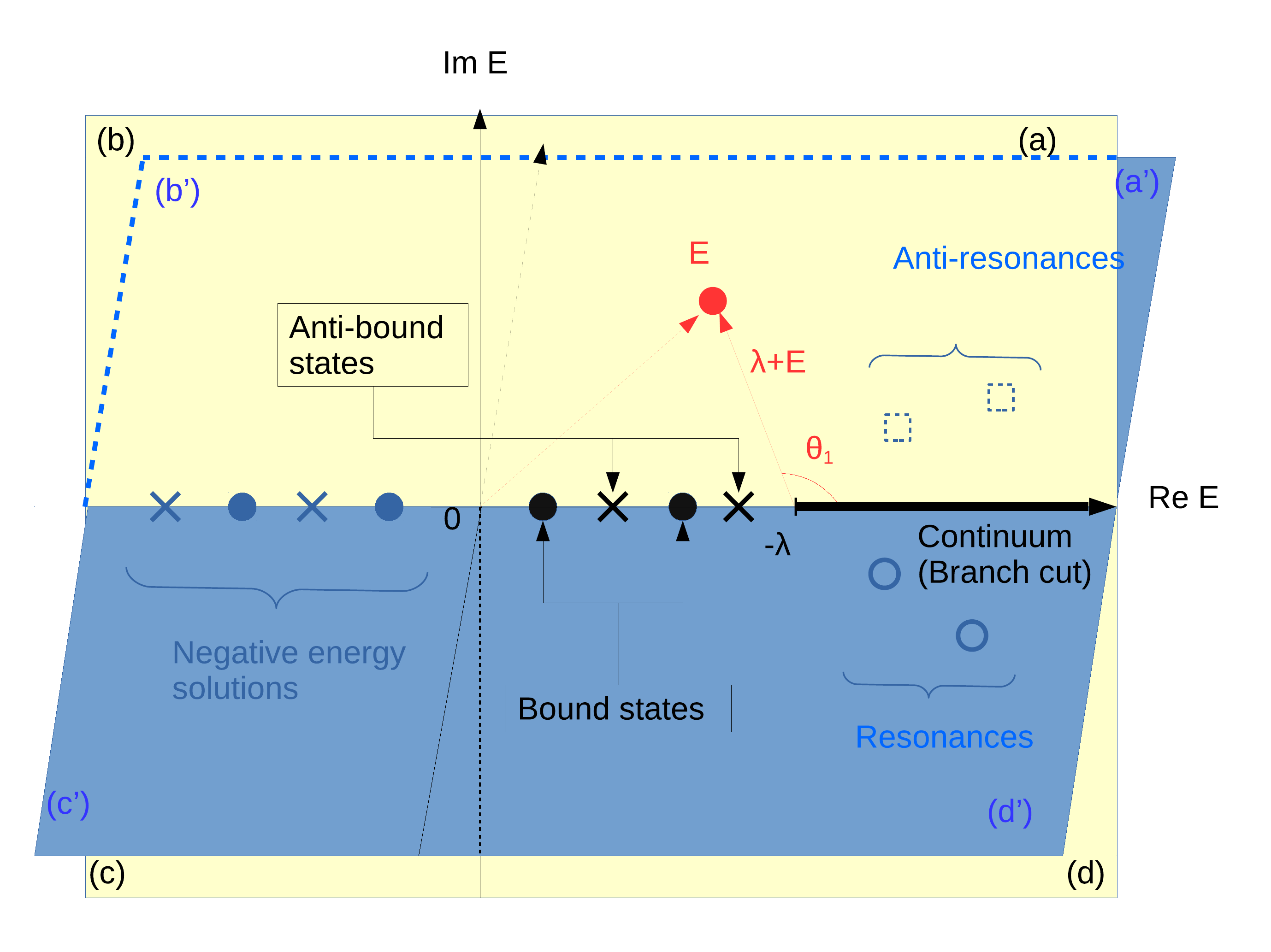}
\caption{(Color online) Complex quasiparticle energy planes with S-matrix poles corresponding bound states,
  anti-bound states, resonances and anti-resonances. The angle $\theta_1$ which is used in
  Eq.(\ref{k1def2}) is also shown. The branch cut corresponding Re $k_1$ line of scattering
  states is shown by the black solid arrow on the real axis above $-\lambda$. Four quadrants
  of the first Riemann sheet (a)-(d) correspond to four areas of the upper half plane of $k_1$
  shown as (a)-(d) in Fig.\ref{fig2}. Four quadrants of the second Riemann sheet (a')-(d')
  correspond to four areas of the lower half plane of $k_1$ shown as (a')-(d') in Fig.\ref{fig2}.}
\label{fig1}
\end{figure}

\begin{figure}[htbp]
\includegraphics[scale=0.25,angle=0]{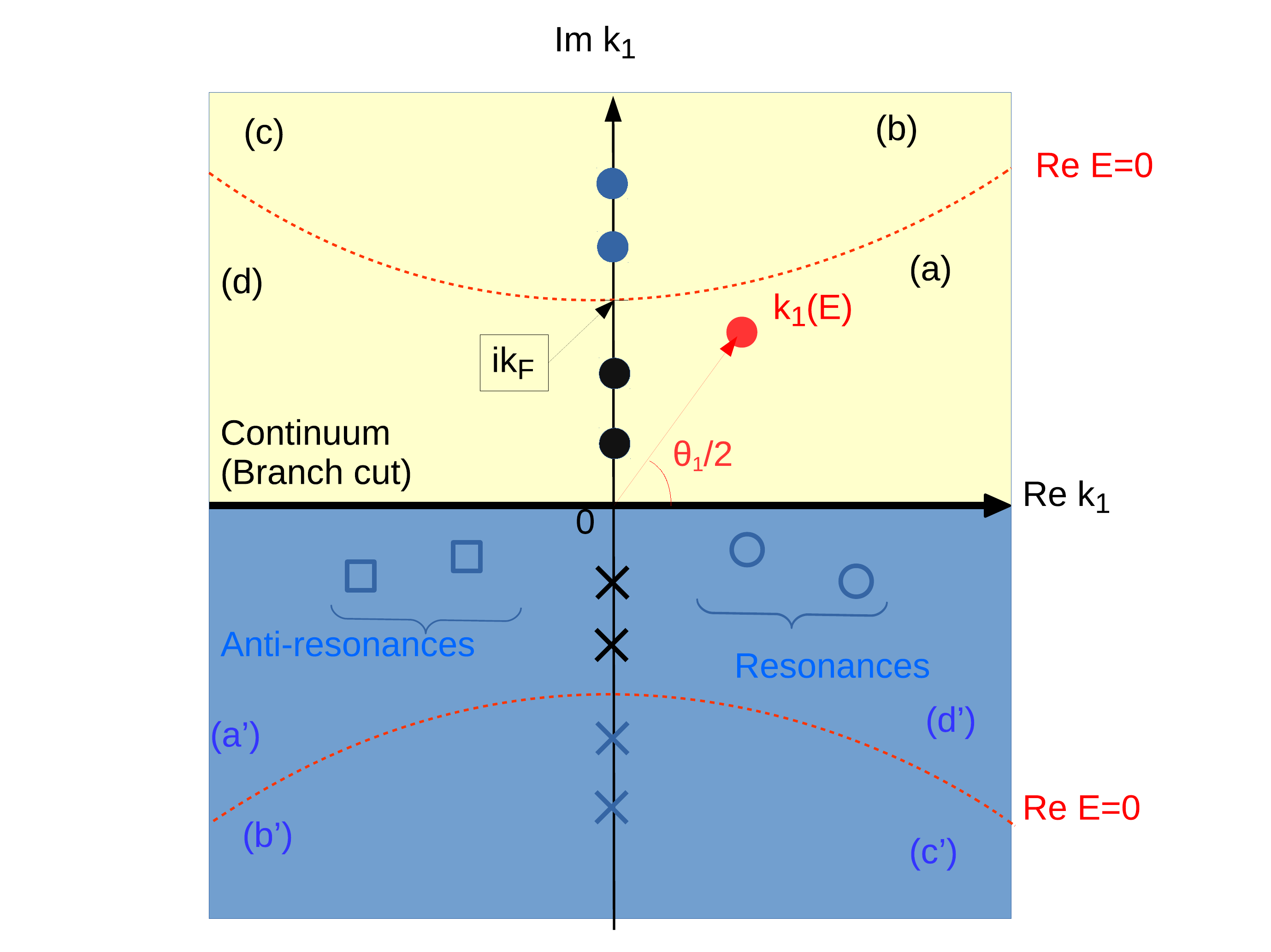}
\caption{(Color online) Complex $k_1$-momentum plane corresponding the complex energy plane shown in
  Fig.\ref{fig1}. Corresponding states are shown by using the same colors and symbols.
  The black solid arrow on Re $k_1$ indicates the scattering states corresponding the
  branch cut on the complex energy plane.
  The upper/lower half plane correspond to the first/second Riemann sheets of the complex
  energy. The hyperbola shown by dotted red curve represent the $k_1$-momentum with
  Re $E=0$. $k_F$ is defined by $k_F=\sqrt{\frac{2m|\lambda|}{\hbar^2}}$.}
\label{fig2}
\end{figure}

The Hartree-Fock-Bogoliubov(HFB) equation
can be represented as
\begin{eqnarray}
  &&
  \left[
    \left(
    \del{^2}{r^2}
    -
    \frac{l(l+1)}{r^2}
    \right)
    \bvec{1}
    -
    \what{S}_z
    \mathcal{U}_{lj}(r)
    \right]
  \phi_{lj}(r;E)
  \nonumber\\
  &&\hspace{2cm}
  =
  -
  \begin{pmatrix}
    k_1^2(E) & 0 \\
    0 & k_2^2(E)
  \end{pmatrix}
  \phi_{lj}(r;E),
  \label{hfbeq}
  \\
  &&
  \mathcal{U}_{lj}(r)
  =
  \frac{2m}{\hbar^2}
  \begin{pmatrix}
    U_{lj}(r) & \Delta(r) \\
    \Delta(r) & -U_{lj}(r)
  \end{pmatrix},
  \label{hfbpot}
  \\
  &&
  \phi_{lj}(r;E)
  =
  \begin{pmatrix}
    \varphi_{1,lj}(r;E)\\
    \varphi_{2,lj}(r;E)
  \end{pmatrix},
  \label{qpwf}
\end{eqnarray}
where $E$ is the quasi-particle energy defined as positive value,
$U_{lj}(r)$ and $\Delta(r)$ are the (Hartree-Fock) mean-field potential and
the pairing potential, respectively.
Where $\bvec{1}$ is the unit matrix, and $\what{S}_z$ is the third component of the Pauli matrix given by
\begin{eqnarray}
  \what{S}_z
  =
  \begin{pmatrix}
    1 & 0 \\
    0 & -1
  \end{pmatrix},
\end{eqnarray}
$k_1(E)$ and $k_2(E)$ are momentum
defined by
\begin{eqnarray}
  k_1(E)&\equiv&\sqrt{\frac{2m}{\hbar^2}(\lambda+E)},
  \label{k1def}
  \\
  k_2(E)&\equiv&\sqrt{\frac{2m}{\hbar^2}(\lambda-E)},
  \label{k2def}
\end{eqnarray}
for upper and lower components of quasiparticle wave functions of HFB equation,
respectively. $\lambda$ is the Fermi energy defined as negative value in
the case of the bound target.
On the complex-$E$ plane, $\lambda\pm E$ can be represented as
\begin{eqnarray}
  \lambda+E&=&|\lambda+E|e^{i\theta_1(E)},
  \label{defE1}
  \\
  \lambda-E&=&|E-\lambda|e^{i(\theta_2(E)+\pi)},
  \label{defE2}
\end{eqnarray}
where $\theta_1(E)$ and $\theta_2(E)$ are angles defined on the complex-$E$ plane
as shown in Figs.\ref{fig1} and \ref{fig2}. The definition range of angle is
$0\leq\theta_{1,2}(E)\leq 2\pi$.

\begin{figure}[htbp]
\includegraphics[scale=0.25,angle=0]{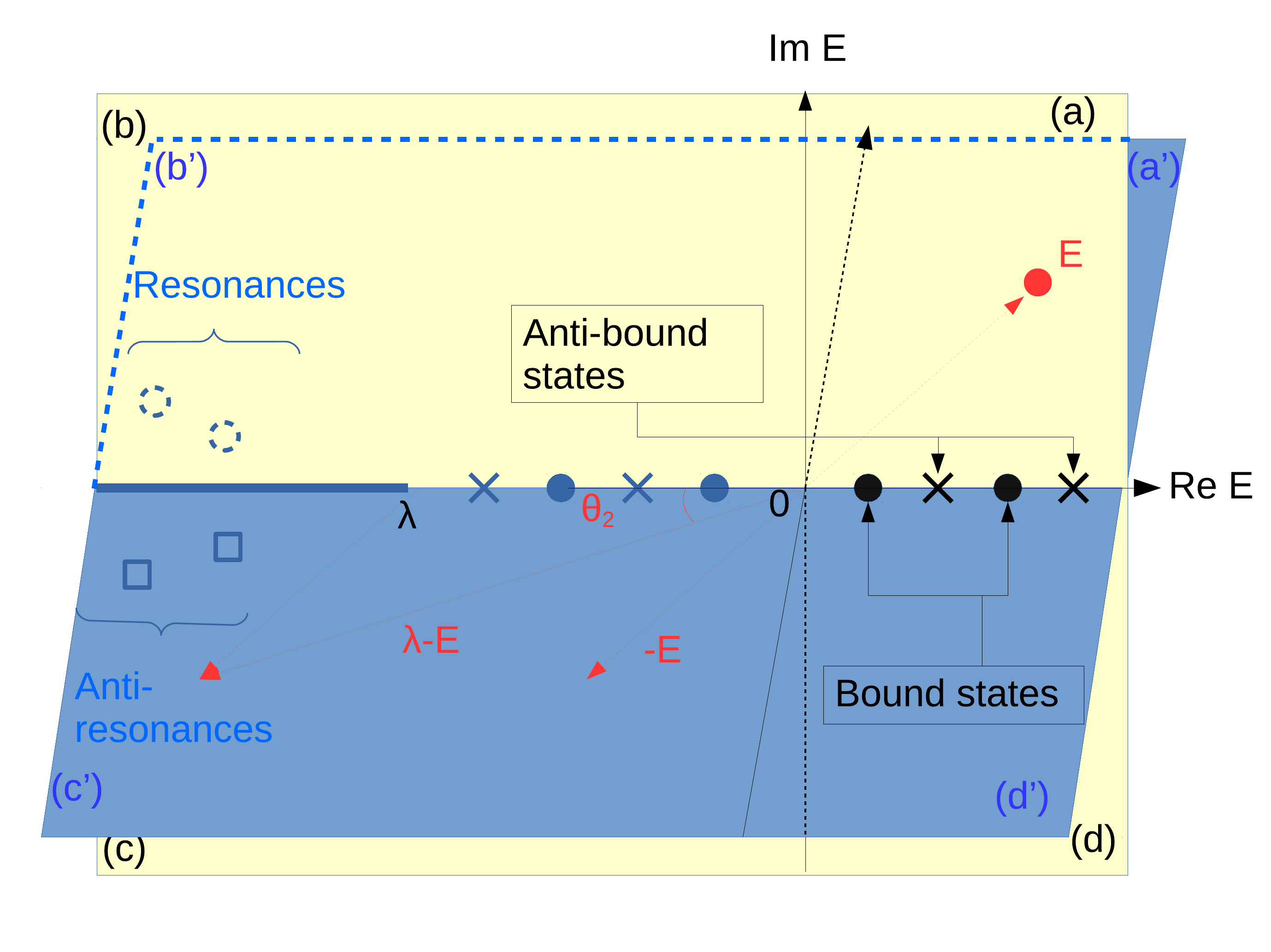}
\caption{(Color online) Same figure with Fig.\ref{fig1} but with branch cut defined by $k_2$-momentum.}
\label{fig3}
\end{figure}

\begin{figure}[htbp]
\includegraphics[scale=0.25,angle=0]{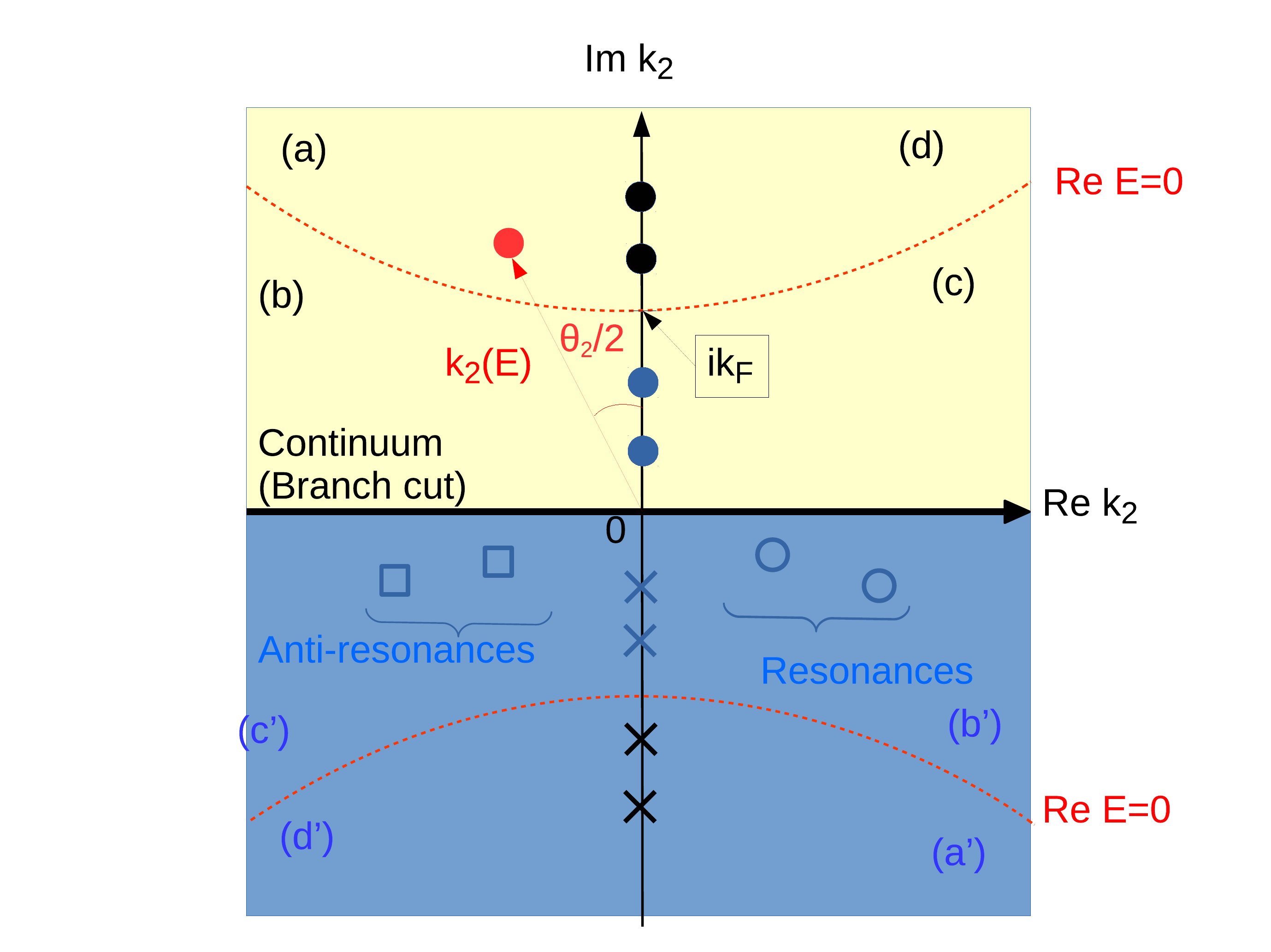}
\caption{(Color online) Same figure with Fig.\ref{fig2} but relevant to $k_2$-momentum.}
\label{fig4}
\end{figure}

Inserting Eqs.(\ref{defE1}) and (\ref{defE2}) into Eqs.(\ref{k1def}) and (\ref{k2def}),
respectively, we can obtain
\begin{eqnarray}
  k_1(E)
  &=&
  |k_1(E)|e^{i\theta_1(E)/2}e^{in\pi}
  \hspace{0.5cm}
  (n=0,1),
  \label{k1def2}
  \\
  |k_1(E)|
  &\equiv&
  \sqrt{\frac{2m}{\hbar^2}|\lambda+E|},
  \label{absk1def}
  \\
  k_2(E)
  &=&
  i|k_2(E)|e^{i\theta_2(E)/2}e^{in\pi}
  \hspace{0.5cm}
  (n=0,1),
  \label{k2def2}
  \\
  |k_2(E)|
  &\equiv&
  \sqrt{\frac{2m}{\hbar^2}|E-\lambda|},
  \label{absk2def},
\end{eqnarray}
the factor $e^{in\pi} (n=0,1)$ is due to the phase ambiguity which shows that
the complex energy plane consists of two Riemann sheets.
By using $n(=0,1)$, let us define the $(n+1)$th Riemann sheet, $E_{n+1}$.
We chose the energy dependent angles $\theta_{1,2}(E)$ have the relation as
\begin{eqnarray}
  &&\theta_{1,2}(E)+2n\pi
  \nonumber\\
  &&\to\theta_{1,2}(E_2)=\theta_{1,2}(E_1)+2\pi,
\end{eqnarray}
so that the first Riemann sheet has the scattering and bound states, and
the second Riemann sheet has the resonance states. Inserting this condition into
Eqs.(\ref{k1def2}) and (\ref{k2def2}), it is easy to find that
$k_{1,2}(E)$ have the relation
\begin{eqnarray}
  k_{1,2}(E_2)=-k_{1,2}(E_1).
\end{eqnarray}
and the upper/lower half of the complex momentum-$k_{1,2}$
correspond to the first/second Riemann sheets respectively.
The branch cut is set for each $k_1(E)$ and $k_2(E)$ as shown in Fig.\ref{fig1},
Fig.\ref{fig2}, Fig.\ref{fig3} and Fig.\ref{fig4}.

\section{Jost function based on the HFB formalism}
\label{sec3}
In general, the double differential equation such as the Schr\"{o}dinger equation
has the regular and irregular solutions. The regular solution can be represented
by the linear combination of the linear independent irregular solutions.
If the irregular solutions satisfy the out-going (or in-coming) boundary condition,
the Jost function is defined as the coefficients for the linear combination~\cite{jost}.

In this section, we derive the Jost function basing on the HFB formalism.
The HFB equation has two independent solutions for regular solutions and irregular
solutions, respectively.


Applying the Green's theorem, it is possible to describe the regular and irregular
solution in the integral equation form.
The Green's theorem for the HFB equation is given by
\begin{eqnarray}
  &&
  \left[
  \chi_{l}^{\mathrm{T}}(r;E)
  \what{S}_z
  \del{\phi_{lj}(r;E)}{r}
  -
  \del{\chi_{l}^{\mathrm{T}}(r;E)}{r}
  \what{S}_z
  \phi_{lj}(r;E)
  \right]_{r=a}^{r=b}
  \nonumber\\
  &&
  =
  \frac{2m}{\hbar^2}
  \int_a^b dr'
  \chi_{l}^{\mathrm{T}}(r';E)
  \mathcal{U}_{lj}(r')
  \phi_{lj}(r';E),
  \label{greentheorem}
\end{eqnarray}
where $\chi_{l}^{\mathrm{T}}(r;E)$ is the free particle wave function
which satisfies
\begin{eqnarray}
  &&
  \left(
  \del{^2}{r^2}
  -
  \frac{l(l+1)}{r^2}
  \right)
  \chi_{l}(r;E)
  =
  -
  \begin{pmatrix}
    k_1^2(E) & 0 \\
    0 & k_2^2(E)
  \end{pmatrix}
  \chi_l(r;E).
  \nonumber\\
  \label{frwv}
\end{eqnarray}
In general, the second derivative equation has two independent solutions;
$\phi_{lj}^{(rs)}(r;E)$ which is regular at the origin $r=0$ and
$\phi_{lj}^{(\pm s)}(r;E)$ satisfies the out-going boundary condition
at the asymptotic limit $r\to\infty$.
The HFB equation has two independent solutions for each $\phi_{lj}^{(rs)}(r;E)$
and $\phi_{lj}^{(\pm s)}(r;E)$ which are described with the subscription $s=1,2$
as introduced in \cite{matsuo}.

By supposing
\begin{eqnarray}
  &&
  \lim_{r\to 0}
  \left[
    \chi_l^{(\pm s) \mathrm{T}}(r;E)
    \what{S}_z
    \del{\phi_l^{(rs')}(r;E)}{r}
    \right.
    \nonumber\\
    &&\hspace{3cm}
    \left.
    -
    \del{\chi_l^{(\pm s)\mathrm{T}}(r;E)}{r}
    \what{S}_z
    \phi_l^{(rs')}(r;E)
    \right]
  \nonumber\\
  &&
  =
  \mp
  \frac{i}{k_{s}(E)}
  \bra s|\what{S}_z|s'\ket ,
  \label{robin}
\end{eqnarray}
with $\chi_l^{(\pm s) \mathrm{T}}(r;E)$ defined by the spherical Hankel function as
\begin{eqnarray}
  \chi_l^{(\pm 1)}(r;E)
  &=&
  \begin{pmatrix}
    rh^{(\pm)}_l(k_1(E)r) \\
    0
  \end{pmatrix},
  \\
  \chi_l^{(\pm 2)}(r;E)
  &=&
  \begin{pmatrix},
    0 \\
    rh^{(\pm)}_l(k_2(E)r)
  \end{pmatrix}
\end{eqnarray}
and taking $a\to 0$ and $b\to r$ for Eq.(\ref{greentheorem}),
we can derive the integral form of $\phi_{lj}^{(rs)}(r;E)$ as
\begin{eqnarray}
  \phi_{lj}^{(rs)}(r;E)
  &=&
  \chi_l^{(rs)}(r;E)
  \nonumber\\
  &&
  +
  \int_0^\infty dr'
  \mathcal{G}_{FR,l}(rr';E)
  \mathcal{U}_{lj}(r')
  \phi_{lj}^{(rs)}(r';E),
  \nonumber\\
  \label{LSrs0}
  \\
  \chi_l^{(rs)}(r;E)
  &=&
  \frac{1}{2}
  \left[
  \chi_l^{(+s)}(r;E)
  +
  \chi_l^{(-s)}(r;E)
  \right],
\end{eqnarray}
with the Green's function defined by
\begin{eqnarray}
  &&
  \mathcal{G}_{FR,l}(rr';E)
  \nonumber\\
  &&=
  -
  \frac{2m}{\hbar^2}
  \sum_{s=1,2}
  \frac{k_s(E)}{2i}
  \theta(r-r')
  \what{S}_z
  \nonumber\\
  &&\times
  \left[
  \chi_l^{(-s)}(r;E)
  \chi_l^{(+s)\mathrm{T}}(r';E)
  -
  \chi_l^{(+s)}(r;E)
  \chi_l^{(-s)\mathrm{T}}(r';E)
  \right]
  \nonumber\\
  \\
  &&=
  \begin{pmatrix}
    g_{FR,l}(rr';k_1(E))
    & 0\\
    0 & -g_{FR,l}(rr';k_2(E))
  \end{pmatrix},
  \label{GFR}
\end{eqnarray}
where
\begin{eqnarray}
  &&
  g_{FR,l}(rr';k)
  \nonumber\\
  &&\equiv
  -
  \frac{2m}{\hbar^2}
  \frac{k}{2i}
  \theta(r-r')rr'
  \nonumber\\
  &&\times
  \left[
    h^{(-)}_l(kr)h^{(+)}_l(kr')
    -
    h^{(-)}_l(kr')h^{(+)}_l(kr)
    \right],
\end{eqnarray}
Note that Eq.(\ref{robin}) is the so-called {\it Robin boundary condition}~\cite{robin} at the origin.
The Robin boundary condition is the combination of the Dirichlet and Neumann
boundary conditions.

The asymptotic boundary condition of $\phi_{lj}^{(\pm s)}(r;E)$
at the limit $r\to\infty$ is given by
\begin{eqnarray}
  \lim_{r\to \infty}
  \phi_{lj}^{(\pm 1)}(r;E)
  &\to&
  \begin{pmatrix}
    rh^{(\pm )}_l(k_1(E) r)\\
    0
  \end{pmatrix},
  \\
  \lim_{r\to \infty}
  \phi_{lj}^{(\pm 2)}(r;E)
  &\to&
  \begin{pmatrix}
    0 \\
    rh^{(\pm )}_l(k_2(E) r)
  \end{pmatrix}.
\end{eqnarray}
Taking $a\to r$ and $b\to\infty$ for Eq.(\ref{greentheorem})
with these boundary conditions, we can obtain the integral form
of $\phi_{lj}^{(\pm s)}(r;E)$ as
\begin{eqnarray}
  &&
  \phi_{lj}^{(\pm s)}(r;E)
  \nonumber\\
  &&
  =
  \chi_l^{(\pm s)}(r;E)
  +
  \int_0^\infty dr'
  \mathcal{G}_{FL,l}(rr';E)
  \mathcal{U}_{lj}(r')
  \phi_{lj}^{(\pm s)}(r';E)
  \nonumber\\
  \label{LSirsol},
\end{eqnarray}
with the Green's function defined by
\begin{eqnarray}
  &&
  \mathcal{G}_{FL,l}(rr';E)
  \nonumber\\
  &&=
  \frac{2m}{\hbar^2}
  \sum_{s=1,2}
  \frac{k_s(E)}{2i}
  \theta(r'-r)
  \what{S}_z
  \nonumber\\
  &&\times
  \left[
  \chi_l^{(-s)}(r;E)
  \chi_l^{(+s)\mathrm{T}}(r';E)
  -
  \chi_l^{(+s)}(r;E)
  \chi_l^{(-s)\mathrm{T}}(r';E)
  \right]
  \nonumber\\
  \\
  &&=
  \begin{pmatrix}
    g_{FL,l}(rr';k_1(E))
    & 0\\
    0 & -g_{FL,l}(rr';k_2(E))
  \end{pmatrix},
  \label{GFL}
\end{eqnarray}
where
\begin{eqnarray}
  &&
  g_{FL,l}(rr';k)
  \nonumber\\
  &&\equiv
  \frac{2m}{\hbar^2}
  \frac{k}{2i}rr'
  \theta(r'-r)
  \nonumber\\
  &&\times
  \left[
    h^{(-)}_l(kr)h^{(+)}_l(kr')
    -
    h^{(-)}_l(kr')h^{(+)}_l(kr)
    \right]
\end{eqnarray}.
It should be noted that we used the value of Wronskian
for the spherical Hankel function given by
\begin{eqnarray}
  W(rh^{(+)}_l(kr),rh^{(-)}_l(kr))=-\frac{2i}{k},
\end{eqnarray}
in order to derive Eq.(\ref{LSirsol}).

By notifying the relation between $\mathcal{G}_{FR,l}$ and $\mathcal{G}_{FL,l}$ given by
\begin{eqnarray}
  &&
  \mathcal{G}_{FR,l}(rr';E)
  =
  \mathcal{G}_{FL,l}(rr';E)
  \nonumber\\
  &&
  -
  \frac{2m}{\hbar^2}
  \sum_{s=1,2}
  \frac{k_s(E)}{2i}
  \what{S}_z
  \nonumber\\
  &&\times
  \left[
  \chi_l^{(-s)}(r;E)
  \chi_l^{(+s)\mathrm{T}}(r';E)
  -
  \chi_l^{(+s)}(r;E)
  \chi_l^{(-s)\mathrm{T}}(r';E)
  \right],
  \nonumber\\
\end{eqnarray}
we can easily find the relation between $\phi_{lj}^{(rs)}$ and $\phi_{lj}^{(\pm s)}$ given by
\begin{eqnarray}
  \phi_{lj}^{(rs)}(r;E)
  &=&
  \frac{1}{2}
  \sum_{s'=1,2}
  \left[
    \left(\mathcal{J}_{lj}^{(+)}(E)\right)_{ss'}
    \phi_{lj}^{(-s')}(r;E)
    \right.
    \nonumber\\
    &&
    \left.
    +
    \left(\mathcal{J}_{lj}^{(-)}(E)\right)_{ss'}
    \phi_{lj}^{(+s')}(r;E)
    \right],
  \label{regiregsol}
\end{eqnarray}
where $\left(\mathcal{J}_{lj}^{(\pm)}(E)\right)_{ss'}$ is the Jost
function defined by
\begin{eqnarray}
  &&
  \left(\mathcal{J}_{lj}^{(\pm)}(E)\right)_{ss'}
  \equiv
  \delta_{ss'}
  \nonumber\\
  &&
  \mp
  \bra s'|\what{S}_z|s'\ket
  \frac{2m}{\hbar^2}
  \frac{k_{s'}(E)}{i}
  \int_0^\infty dr
  \chi_l^{(\pm s')\mathrm{T}}(r;E)
  \mathcal{U}_{lj}(r)
  \phi_{lj}^{(rs)}(r;E).
  \nonumber\\
  \label{jost1}
\end{eqnarray}
This Jost function can be also expressed as
\begin{eqnarray}
  &&
  \left(\mathcal{J}_{lj}^{(\pm)}(E)\right)_{ss'}
  \nonumber\\
  &&=
  \delta_{ss'}
  \nonumber\\
  &&
  \mp
  \bra s'|\what{S}_z|s'\ket
  \frac{k_{s'}(E)}{i}
  \frac{2m}{\hbar^2}
  \int_0^\infty dr
  \chi_{l}^{(rs)\mathrm{T}}(r;E)
  \mathcal{U}_{lj}(r)
  \phi_{lj}^{(\pm s')}(r;E)
  \nonumber\\
  \label{jost2}
  \\
  &&=
  \pm
  \bra s'|\what{S}_z|s'\ket
  \frac{k_{s'}(E)}{i}
  \frac{2m}{\hbar^2}
  w_{lj}^{s,\pm s'}(E),
  \label{jost3}
\end{eqnarray}
where $w_{lj}^{s,\pm s'}(E)$ is the Wronskian for the
HFB solutions defined by
\begin{eqnarray}
  &&
  w_{lj}^{s,\pm s'}(E)
  \equiv
  \frac{\hbar^2}{2m}
  \left[
  \del{\phi_l^{(\pm s') \mathrm{T}}(r;E)}{r}
  \what{S}_z
  \phi_l^{(rs)}(r;E)
  \right.
  \nonumber\\
  &&\hspace{2cm}
  -
  \left.
  \phi_l^{(\pm s') \mathrm{T}}(r;E)
  \what{S}_z
  \del{\phi_l^{(rs)}(r;E)}{r}
  \right],
\end{eqnarray}
as introduced in \cite{matsuo,fayans,belyaev}.
\begin{figure}[htbp]
\includegraphics[scale=0.75,angle=0]{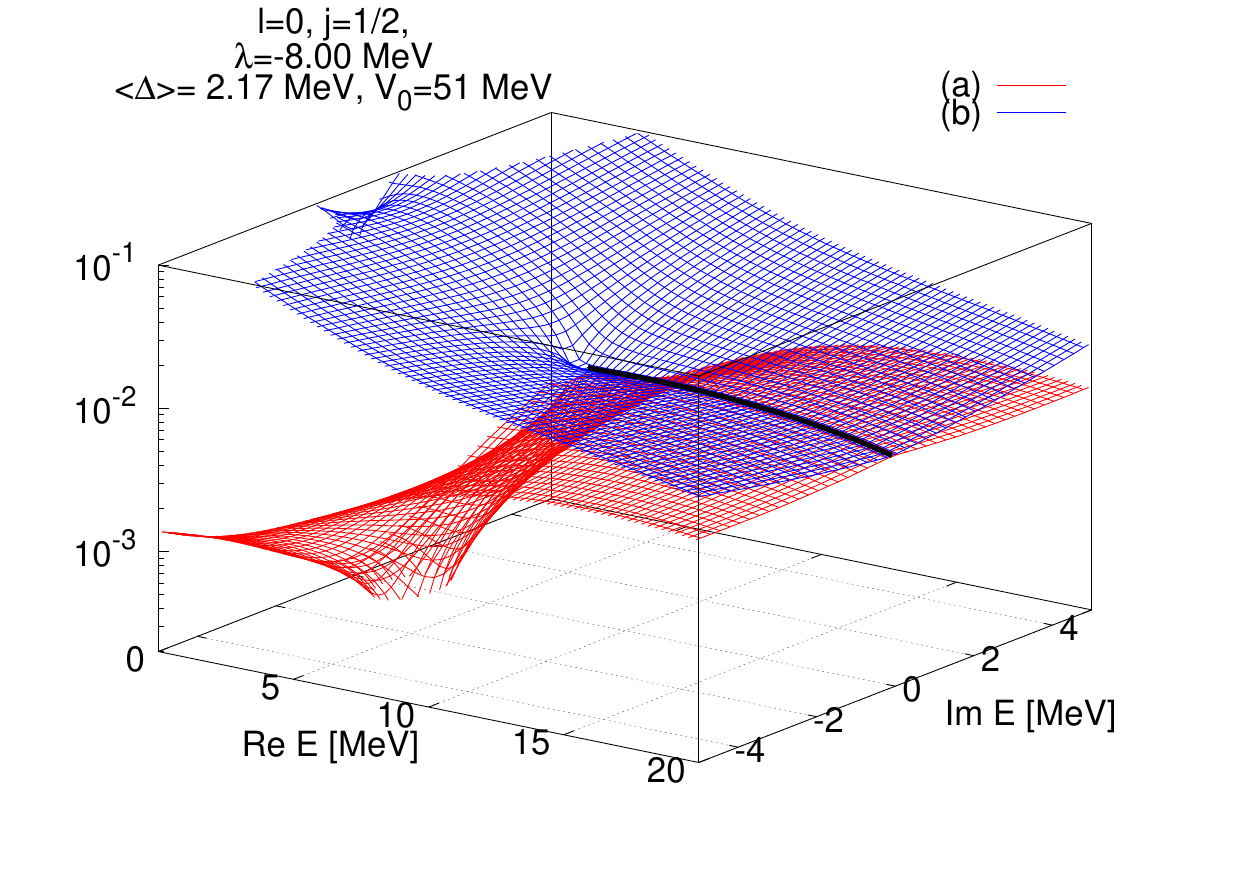}
\caption{(Color online) The absolute value surface of the denominator
  ((a) red colored surface) and the denominator ((b) blue colored surface)
  of the S-matrix Eq.(\ref{S11}) for $s_{1/2}$. The black solid curve is a
  curve where the red and blue surfaces ((a) and (b)) are touched each other.}
\label{fig5}
\end{figure}
Note that the off-diagonal components of the Jost function
becomes $\left(\mathcal{J}_{lj}^{(\pm)}(E)\right)_{12}=\left(\mathcal{J}_{lj}^{(\pm)}(E)\right)_{21}=0$
at the no pairing limit $\bra\Delta\ket\to 0$.

The asymptotic behaviour at $r\to\infty$ of Eq.(\ref{regiregsol}) is given by
\begin{eqnarray}
  &&
  \lim_{r\to\infty}
  \phi_{lj}^{(rs)}(r;E)
  \nonumber\\
  &&\sim
  \frac{i^{l+1}}{2k_1(E)}
  \left[
    \left(\mathcal{J}_{lj}^{(+)}(E)\right)_{s1}
    \begin{pmatrix}
      e^{-ik_1(E)r} \\
      0
    \end{pmatrix}
    \right.
    \nonumber\\
    &&
    \hspace{1.5cm}
    \left.
    +
    (-)^{l+1}
    \left(\mathcal{J}_{lj}^{(-)}(E)\right)_{s1}
    \begin{pmatrix}
      e^{+ik_1(E)r} \\
      0
    \end{pmatrix}
    \right]
  \nonumber\\
  &&+
  \frac{i^{l+1}}{2k_2(E)}
  \left[
    \left(\mathcal{J}_{lj}^{(+)}(E)\right)_{s2}
    \begin{pmatrix}
      0 \\
      e^{-ik_2(E)r}
    \end{pmatrix}
    \right.
    \nonumber\\
    &&
    \hspace{1.5cm}
    \left.
    +
    (-)^{l+1}
    \left(\mathcal{J}_{lj}^{(-)}(E)\right)_{s2}
    \begin{pmatrix}
      0 \\
      e^{+ik_2(E)r}
    \end{pmatrix}
    \right].
\end{eqnarray}
If the the complex quasiparticle energy $E$ belongs to the 1st Riemann sheet ($E=E_1$),
the complex momentum $k_1$ and $k_2$ belong to the upper-half plane of each
momentum planes as explained in Sec.\ref{sec2}.
The condition for the bound states are, therefore, given by
\begin{eqnarray}
  \left(\mathcal{J}_{lj}^{(+)}(E_1)\right)_{s1}
  =
  \left(\mathcal{J}_{lj}^{(+)}(E_1)\right)_{s2}
  =0,
\end{eqnarray}
or
\begin{eqnarray}
  \det \left(\mathcal{J}_{lj}^{(+)}(E_1)\right)=0.
\end{eqnarray}
When $\det \left(\mathcal{J}_{lj}^{(+)}(E)\right)\neq 0$,
the scattering wave function $\psi_{lj}^{(+)}(r;E)$ can be defined
using the inverse of the Jost function $\left(\mathcal{J}_{lj}^{(+)}(E)\right)^{-1}$ as
\begin{eqnarray}
  \psi_{lj}^{(+)}(r;E)
  \equiv
  \sum_{s=1,2}
  \left(\mathcal{J}_{lj}^{(+)}(E)\right)^{-1}_{1s}
  \phi_{lj}^{(rs)}(r;E)
  \label{defpsi}.
\end{eqnarray}
since the upper component of the HFB solution corresponding
to particle states at no pairing limit ($\Delta(r)\to 0$)
in the positive energy region (Re $E>0$).

The asymptotic behaviour of $\psi_{lj}^{(+)}(r;E)$ is given by
\begin{eqnarray}
  &&
  \lim_{r\to\infty}
  \psi_{lj}^{(+)}(r;E)
  \nonumber\\
  &&\sim
  \frac{i^{l+1}}{2k_1(E)}
  \left[
    \begin{pmatrix}
      e^{-ik_1(E)r} \\
      0
    \end{pmatrix}
    \right.
    \nonumber\\
    &&
    \hspace{2.5cm}
    \left.
    +
    (-)^{l+1}
    S^{11}_{lj}(E)
    \begin{pmatrix}
      e^{+ik_1(E)r} \\
      0
    \end{pmatrix}
    \right]
  \nonumber\\
  &&\hspace{0.5cm}
  +
  \frac{(-i)^{l+1}}{2k_2(E)}
  S^{12}_{lj}(E)
  \begin{pmatrix}
    0 \\
    e^{+ik_2(E)r}
  \end{pmatrix},
  \label{asympsi}
\end{eqnarray}
where
\begin{eqnarray}
  S^{ss'}_{lj}(E)
  \equiv
  \sum_{s''=1,2}
  \left(\mathcal{J}_{lj}^{(+)}(E)\right)^{-1}_{ss''}
  \left(\mathcal{J}_{lj}^{(-)}(E)\right)_{s''s'}.
  \label{defSmat}
\end{eqnarray}
Since $k_1$ becomes real, and $k_2$ becomes pure imaginary on the upper-half
plane of each momentum as seen in Fig.\ref{fig1}-\ref{fig4} on the scattering
states (the real axis of $E$, Re $E > -\lambda (>0)$), the second term is
vanished at $r\to\infty$ in the r.h.s of Eq.(\ref{asympsi}). And $S_{lj}^{11}(E)$
can be interpreted as the S-matrix for the nucleon scattering on open-shell
nucleus within the HFB formalism.

\begin{widetext}
According to the definition of $S^{ss'}_{lj}(E)$ (Eq.(\ref{defSmat})),
$S_{lj}^{11}(E)$ can be expressed as
\begin{eqnarray}
  S_{lj}^{11}(E)
  =
  \frac{
    \left(\mathcal{J}_{lj}^{(+)}(E)\right)_{22}
    \left(\mathcal{J}_{lj}^{(-)}(E)\right)_{11}
    -
    \left(\mathcal{J}_{lj}^{(+)}(E)\right)_{12}
    \left(\mathcal{J}_{lj}^{(-)}(E)\right)_{21}
  }
       {
         \left(\mathcal{J}_{lj}^{(+)}(E)\right)_{22}
         \left(\mathcal{J}_{lj}^{(+)}(E)\right)_{11}
         -
         \left(\mathcal{J}_{lj}^{(+)}(E)\right)_{12}
         \left(\mathcal{J}_{lj}^{(+)}(E)\right)_{21}
       }.
       \label{S11}
\end{eqnarray}
\end{widetext}

If the quasiparticle energy $E$ moves from the 1st quadrant to the 4th quadrant
by passing through branch cut corresponding to $k_1$ ($E\to E^*$ from (a) to (d')
in Fig.\ref{fig1}), $k_1(E^*)$ is given by
\begin{eqnarray}
  k_1(E^*)=k_1^*(E),
  \label{k1ch}
\end{eqnarray}
which is corresponding to the motion from (a) to (d') in Fig.\ref{fig2}.

Since there is no branch cut for $k_2$ between the 1st and 4th quadrants,
\begin{eqnarray}
  k_2(E^*)=-k_2^*(E).
  \label{k2ch}
\end{eqnarray}
(corresponding the motion from (a) to (d) in Fig.\ref{fig4}).

By applying Eqs.(\ref{k1ch}) and (\ref{k2ch}) to Eqs.(\ref{LSrs0})
and (\ref{LSirsol}), we find
\begin{eqnarray}
  \phi_{lj}^{(r1)}(r;E^*)
  &=&
  \phi_{lj}^{(r1)*}(r;E),
  \\
  \phi_{lj}^{(r2)}(r;E^*)
  &=&
  (-)^l
  \phi_{lj}^{(r2)*}(r;E),
  \\
  \phi_{lj}^{(\pm 1)}(r;E^*)
  &=&
  \phi_{lj}^{(\mp 1)*}(r;E),
  \\
  \phi_{lj}^{(\pm 2)}(r;E^*)
  &=&
  (-)^l
  \phi_{lj}^{(\pm 2)*}(r;E),
\end{eqnarray}
and the Jost function as a function of $E^*$ have the
following properties as
\begin{eqnarray}
  \left(\mathcal{J}_{lj}^{(\pm)}(E^*)\right)_{11}
  &=&
  \left(\mathcal{J}_{lj}^{(\mp)*}(E)\right)_{11},
  \\
  \left(\mathcal{J}_{lj}^{(\pm)}(E^*)\right)_{12}
  &=&
  (-)^l
  \left(\mathcal{J}_{lj}^{(\pm)*}(E)\right)_{12},
  \\
  \left(\mathcal{J}_{lj}^{(\pm)}(E^*)\right)_{21}
  &=&
  (-)^l
  \left(\mathcal{J}_{lj}^{(\mp)*}(E)\right)_{21},
  \\
  \left(\mathcal{J}_{lj}^{(\pm)}(E^*)\right)_{22}
  &=&
  \left(\mathcal{J}_{lj}^{(\pm)*}(E)\right)_{22},
\end{eqnarray}
Thus, we finally can find that $S_{lj}^{11}$ can be represented by
\begin{widetext}
  \begin{eqnarray}
  S_{lj}^{11}(E^*)
  &&=
  \frac{
    \left(\mathcal{J}_{lj}^{(+)*}(E)\right)_{22}
    \left(\mathcal{J}_{lj}^{(+)*}(E)\right)_{11}
    -
    \left(\mathcal{J}_{lj}^{(+)*}(E)\right)_{12}
    \left(\mathcal{J}_{lj}^{(+)*}(E)\right)_{21}
  }
       {
         \left(\mathcal{J}_{lj}^{(+)*}(E)\right)_{22}
         \left(\mathcal{J}_{lj}^{(-)*}(E)\right)_{11}
         -
         \left(\mathcal{J}_{lj}^{(+)*}(E)\right)_{12}
         \left(\mathcal{J}_{lj}^{(-)*}(E)\right)_{21}
       }
       =
       1/S_{lj}^{11 *}(E),
       \label{S11-2}
  \end{eqnarray}
\end{widetext}
when $E^*$ belongs to the 2nd Riemann sheet defined by
$k_1$ in the positive energy region (Re $E>0$).
From this result, we can see that zero points of
the denominator and numerator of Eq.(\ref{S11})
represent the S-matrix poles on the 1st and 2nd
Riemann sheet of the complex energy plane,
respectively.

Also it is obvious that
\begin{eqnarray}
  |S_{lj}^{11}(E)|^2=1,
  \label{unitality}
\end{eqnarray}
when $E=E^*$ in Eq.(\ref{S11-2}). This shows that
$S_{lj}^{11}(E)$ satisfies the Unitarity of the S-matrix
if $E$ stands on the branch cut for the positive energy
region.

The denominator of the S-matrix Eq.(\ref{S11}) is equal to the
determinant of the Jost function.
Since the following properties of the Jost function
\begin{eqnarray}
  \left(\mathcal{J}_{lj}^{(\pm)}(E)\right)_{22}
  &=&
  \left(\mathcal{J}_{lj}^{(\pm)}(-E)\right)_{11},
  \\
  \left(\mathcal{J}_{lj}^{(\pm)}(E)\right)_{12}
  &=&
  -
  \left(\mathcal{J}_{lj}^{(\pm)}(-E)\right)_{21},
\end{eqnarray}
can be found by applying $k_2(-E)=k_1(E)$ to the definition of the Jost function (Eqs.(\ref{jost1})-(\ref{jost3})),
we can find the symmetric property of the determinant of the Jost function as
\begin{eqnarray}
  \det\left(\mathcal{J}_{lj}^{(\pm)}(-E)\right)
  =\det\left(\mathcal{J}_{lj}^{(\pm)}(E)\right).
\end{eqnarray}
Therefore, the zeros of the denominator are given symmetrically to the imaginary axis of the complex energy (Im $E$),
and are corresponding the eigen solutions of the HFB Hamiltonian. The S-matrix poles given by zeros of
the numerator represent the resonances.

In Fig.\ref{fig5}, we show the absolute value surfaces for the denominator
and the numerator of the S-matrix (Eq.(\ref{S11}) for $s_{1/2}$ which is
calculated numerically using the parameter sets and Woods-Saxon potential
introduced in Sec.\ref{sec-num} adopting $\lambda=-8$ MeV and $\bra\Delta\ket=2.17$
MeV for the Fermi energy and the mean pairing gap, respectively.
The denominator and numerator of the S-matrix are touched along the branch cut
defined in the positive energy region.

\begin{figure}[htbp]
\includegraphics[scale=0.7,angle=0]{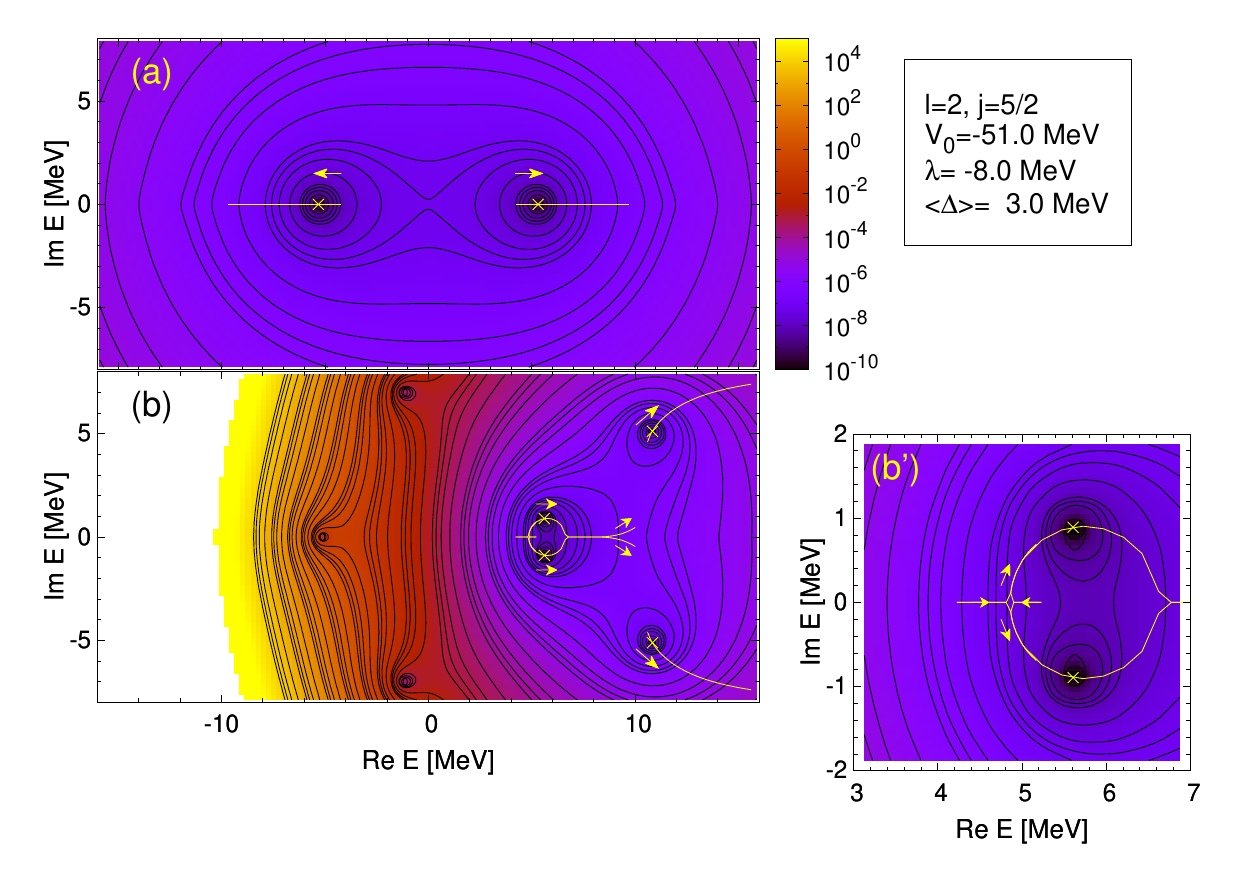}
\caption{(Color online) The square of the denominator (panel (a)) and the numerator (panel (b))
  of the S-matrix (Eq.(\ref{S11})) for $d_{5/2}$ with $\bra\Delta\ket=3.0$ MeV and $\lambda=-8.0$ MeV
  are shown on the complex quasiparticle energy plane. The yellow \text{\sffamily X}-symbols show the position
  of zeros which represent the poles of the S-matrix. The yellow curves is the trajectories of zeros
  obtained by varying the pairing gap $\bra\Delta\ket$ from 0.0 to 10.0 MeV. The arrows represent the direction
  of \text{\sffamily X}.}
\label{fig6-1}
\end{figure}

The T-matrix for the neutron elastic scattering on
open-shell nucleus can be derived as
\begin{eqnarray}
  &&
  T_{lj}(E)
  \equiv
  \frac{i}{2}
  \left(
  S_{lj}^{11}(E)-1
  \right)
  \nonumber\\
  &&=
  \frac{i}{2}
  \sum_{s=1,2}
  \left(\mathcal{J}_{lj}^{(+)}(E)\right)^{-1}_{1s}
  \left[
    \left(\mathcal{J}_{lj}^{(-)}(E)\right)_{s1}
    -
    \left(\mathcal{J}_{lj}^{(+)}(E)\right)_{s1}
    \right]
  \nonumber\\
  &&=
  \frac{2m k_{1}(E)}{\hbar^2}
  \int_0^\infty dr
  \chi_l^{(r1)\mathrm{T}}(r;E)
  \mathcal{U}_{lj}(r)
  \psi_{lj}^{(+)}(r;E)
  \label{tmat}.
\end{eqnarray}
Note that Eqs.(\ref{jost1}) and (\ref{defpsi}) was used
to derive Eq.(\ref{tmat}).

When the S-matrix satisfies the Unitarity (Eq.(\ref{unitality}), the T-matrix
satisfies the optical theorem given by
\begin{eqnarray}
  -\mbox{ Im } T_{lj}(E)=|T_{lj}(E)|^2,
\end{eqnarray}
where $E$ is supposed to be existing on the branch cut.
The total cross section for nucleon scattering $\sigma_{lj}(\hbar\omega)$ is given by
\begin{eqnarray}
  \sigma(E_{i})
  &=&
  \sum_{lj}
  \sigma_{lj}(E_{i}),
  \label{xsectot}
  \\
  \sigma_{lj}(E_{i})
  &=&
  \frac{\hbar^2}{2m}
  \frac{\pi}{E_{i}}
  \frac{2j+1}{2}
  |T_{lj}(E_{i}-\lambda)|^2,
  \label{xsecpart}
\end{eqnarray}
where $E_{i}$ is the incident nucleon energy defied by
\begin{eqnarray}
  E_{i}=E+\lambda.
  \label{ei}
\end{eqnarray}
\begin{figure}[htbp]
\includegraphics[scale=0.45,angle=0]{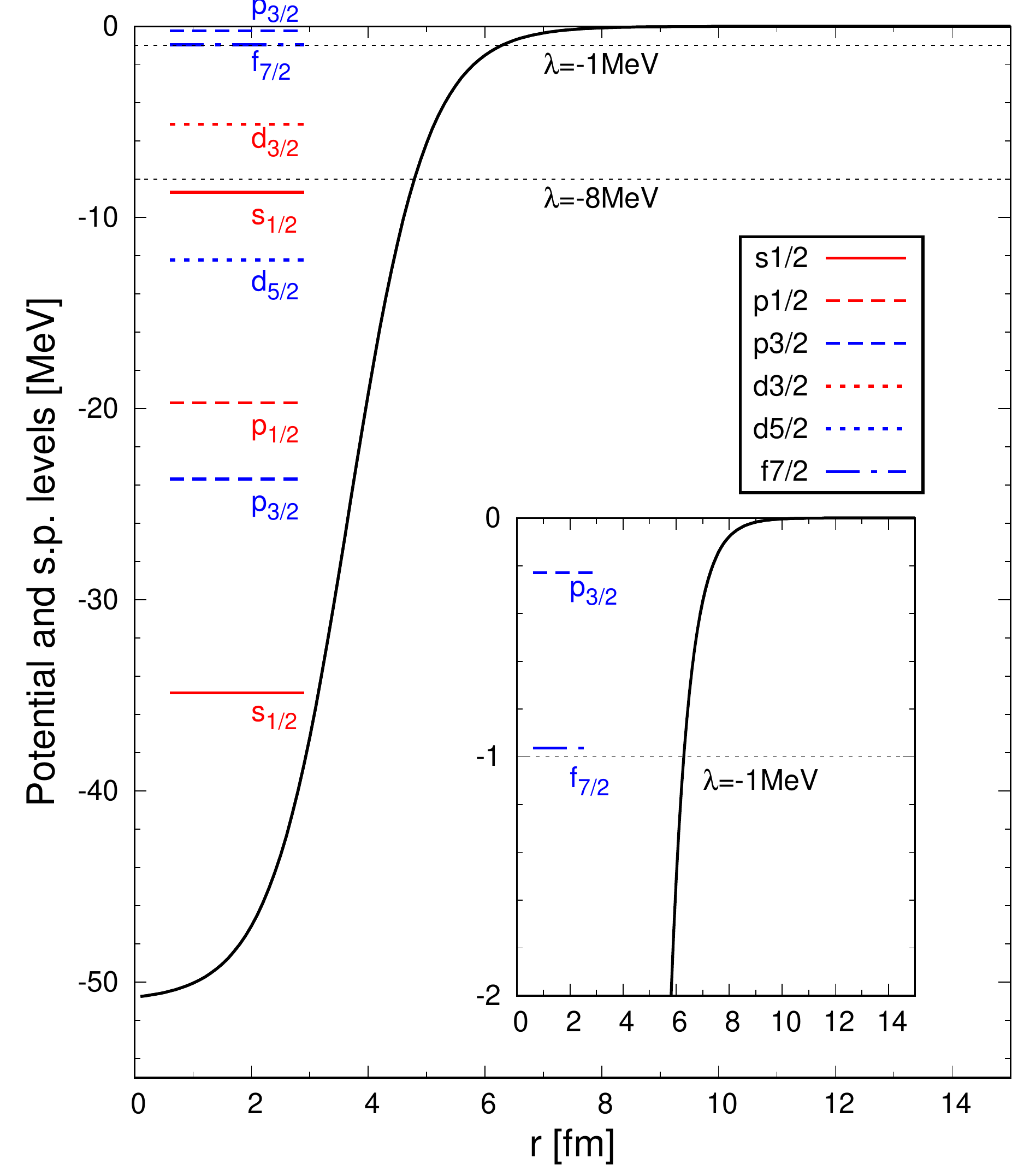}
\caption{(Color online) The Woods-Saxon potential and the single-particle levels obtained in the
  no pairing limit. The lines of Fermi energy $\lambda=-1$ and $-8$ MeV are displayed
  together. }
\label{fig0}
\end{figure}

\section{Numerical results}
\label{sec-num}
In this study, we adopt the Woods-Saxon potential which is given by
\begin{eqnarray}
  U_{lj}(r)
  &=&
  \left[
    V_0
    +
    V_{ls}
    \bvec{l}\idot\bvec{s}
    \frac{r_0^2}{r}
    \frac{d}{dr}
    \right]
  f(r),
  \\
  f(r)
  &=&
  \frac{1}{1+\exp\left(\frac{r-R}{a}\right)},
\end{eqnarray}
in the standard form for the Hartree-Fock
mean field $U_{lj}(r)$ in Eq.(\ref{hfbeq})
with use of the standard parameters given
by \cite{hamamoto,bohr}. The mass number $A=24$ is
used for the parameter given by $R=r_0A^{\frac{1}{3}}$.

The volume-type pairing field $\Delta(r)$ described by
\begin{eqnarray}
  \Delta(r)
  =
  V_{pair}f(r)
\end{eqnarray}
is adopted in the paper.

We define the average pairing gap $\bra\Delta\ket$ by
\begin{eqnarray}
  \bra\Delta\ket
  \equiv
  \frac{\int r^2dr \Delta(r) f(r)}{\int r^2dr f(r)},
\end{eqnarray}
as introduced in \cite{hamamoto}.
We solve the integral equation Eq.(\ref{LSirsol}) numerically, and obtain
$\phi_{lj}^{(\pm s)}(r;E) (s=1,2)$ in order to calculate the Jost function
with Eq.(\ref{jost2}) in this study. It should be noted that we confirmed that
the Jost function by Eqs.(\ref{jost1})-(\ref{jost3}) give the
same numerical results.

The square of the denominator (panel (a)) and the numerator (panel (b))
of the S-matrix (Eq.(\ref{S11})) for $d_{5/2}$ with $\bra\Delta\ket=3.0$ MeV and $\lambda=-8.0$ MeV
are shown on the complex quasiparticle energy plane in Fig.\ref{fig6-1}. The yellow colored \text{\sffamily X}-symbols
are zeros of the plotted function determined by the Nelder-Mead method~\cite{nelder}.
The yellow curves are the trajectories of the S-matrix poles with varying $\bra\Delta\ket$ from 0.0 to 10.0 MeV.
The arrows represent the direction of motion of zeros indicated by \text{\sffamily X}-symbols.

\begin{figure}[htbp]
\includegraphics[scale=0.7,angle=0]{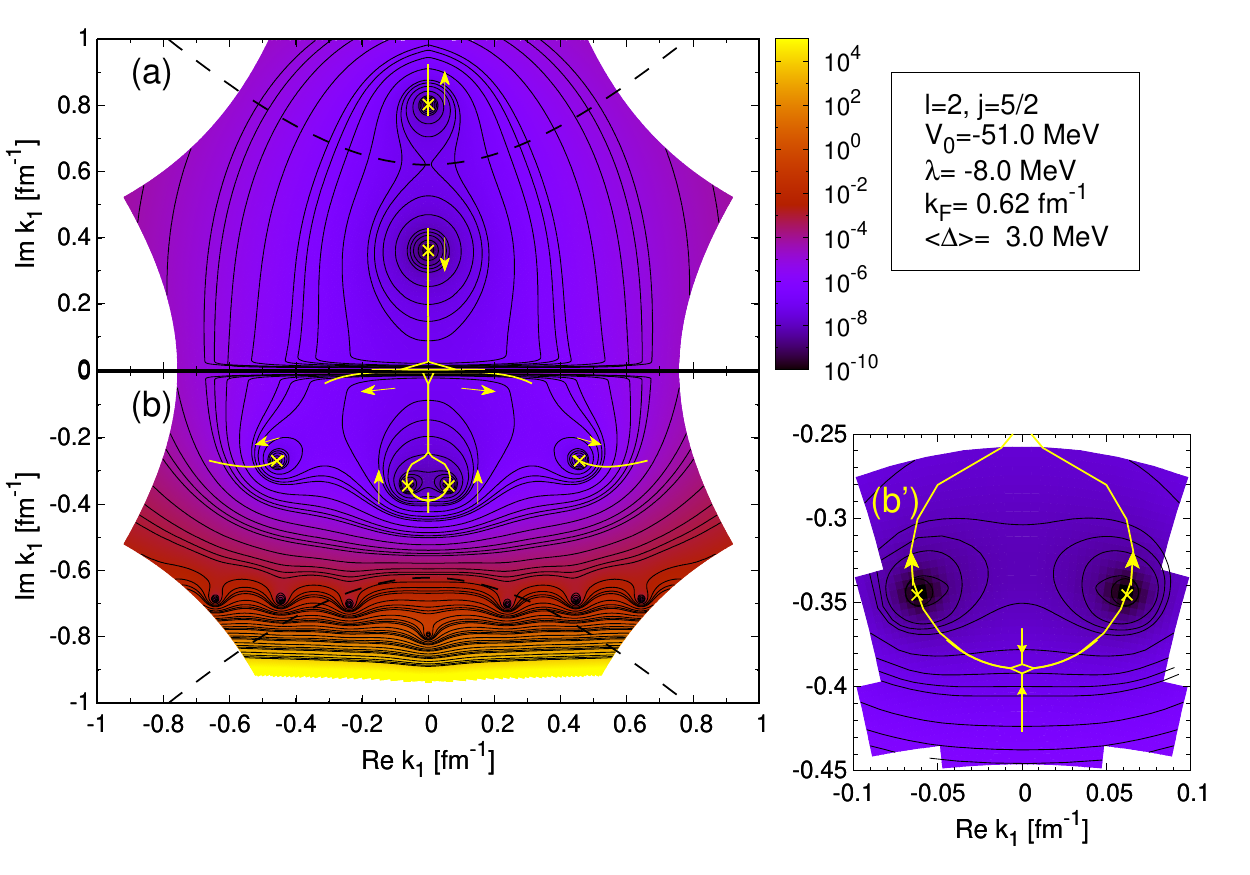}
\caption{(Color online) The same figure with Fig.\ref{fig6-1} but on the complex $k_1$-momentum plane. The upper- and lower-half plane
  ((a) and (b)) correspond to the panels (a) and (b) in Fig.\ref{fig6-1}, respectively. The dashed curves
  show curves where Re $E =0$ MeV. }
\label{fig6-2}
\end{figure}

The same results are shown on the complex $k_1$-plane in Fig.\ref{fig6-2}. Panels (a) and (b) in Fig.\ref{fig6-1}
correspond to the (a)upper and (b)lower figures in Fig.\ref{fig6-2}, respectively. The dashed curves seen in the
upper-half and lower-half plane represent Re $E =0$ which pass through the points at $k_1 = \pm i k_F$. $k_F$ is the Fermi
momentum defined by $k_F\equiv\sqrt{\frac{2m|\lambda|}{\hbar^2}}$. The above/below regions on upper/lower panels
is corresponding to the negative energy regions (left-half planes) of 1st/2nd Riemann sheets, respectively,
as explained in Figs.\ref{fig1} and \ref{fig2}.

The S-matrix poles on the 1st Riemann sheet are existing on the real axis, and have a tendency to
go away from $E=0$ along the real axis. Corresponding poles on $k_1$-momentum plane are existing on the
imaginary axis and go away from the Fermi momentum along the imaginary axis as the pairing gap is
increased.

\begin{figure}[htbp]
\includegraphics[scale=0.45,angle=0]{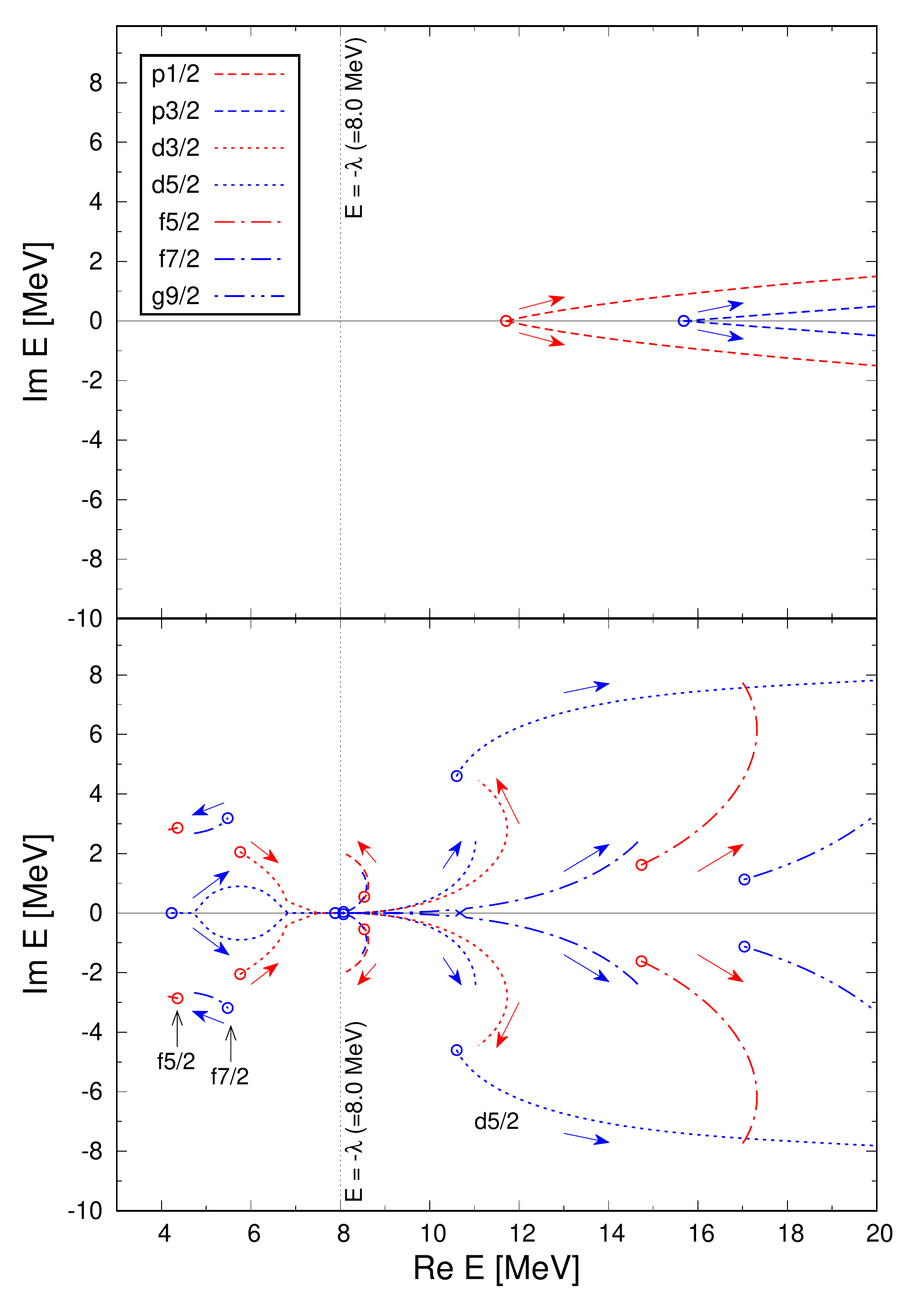}
\caption{(Color online) Trajectories of the S-matrix poles on the 2nd Riemann sheet
  with varying the pairing gap $\bra\Delta\ket$ from 0 to 10 MeV. \text{\sffamily o}-symbols represent
  the position of poles at the no pairing limit ($\bra\Delta\ket=0$ MeV).
  The hole-like and particle-like poles are classified in the upper and lower panels,
  respectively ({\it see text}).
}
\label{fig7-1}
\end{figure}
\begin{figure}[htbp]
\includegraphics[scale=0.4,angle=0]{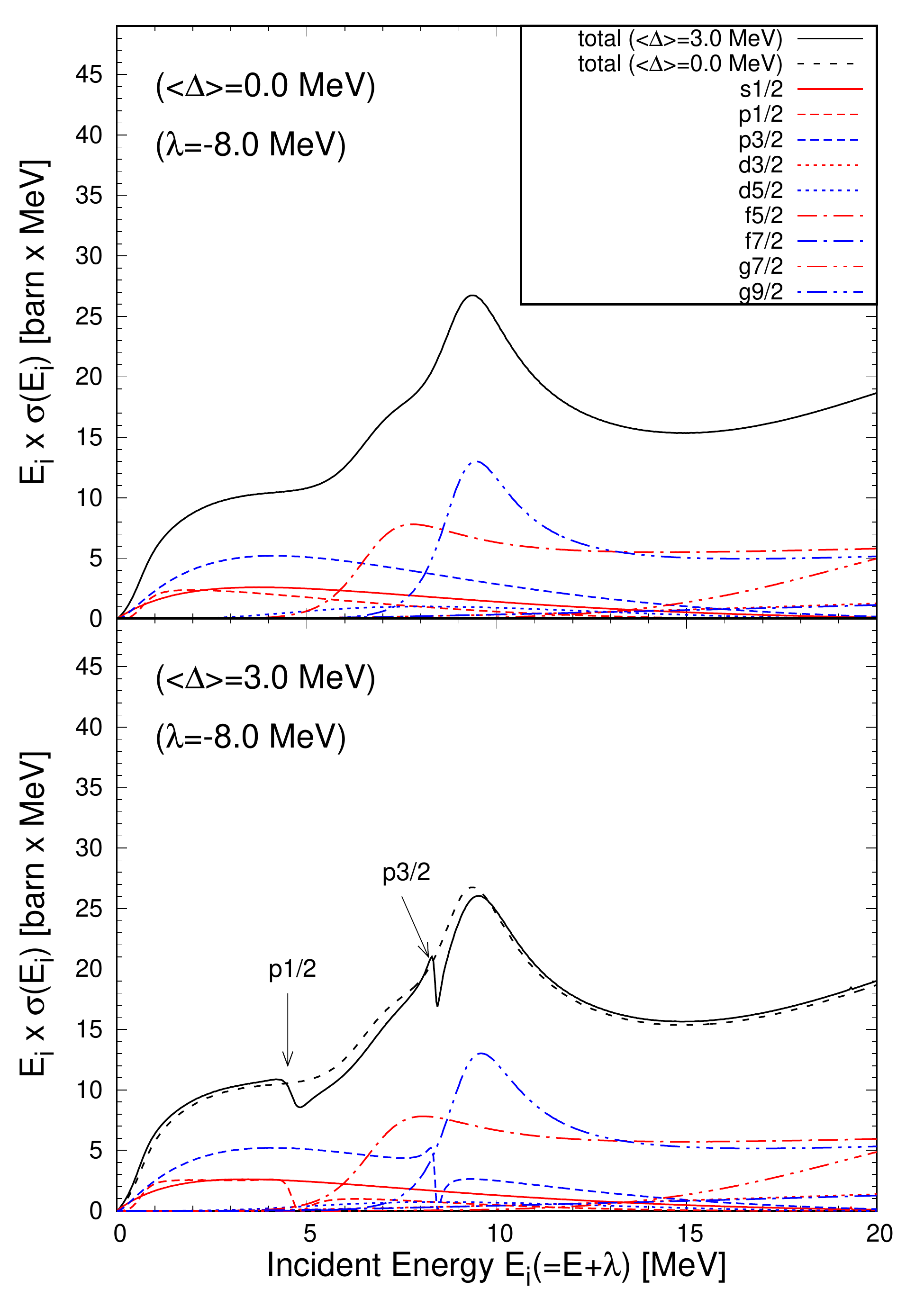}
\caption{(Color online) The total cross section and the partial cross sections for the neutron elastic scattering with $\lambda=-8.0$ MeV.}
\label{fig10}
\end{figure}

The S-matrix poles existing on the real axis of the 2nd Riemann sheet move along the real axis
by taking a circular trajectory as the pairing gap is increased.
It seems to avoid a certain point existing around $E=-\frac{3}{4}\lambda=6$ MeV shown as (b') in
Fig.\ref{fig6-1}. The certain point corresponds to $k_1=-ik_F/2$ on the complex $k_1$-momentum
plane. The poles become resonances at $E>-\lambda$ with the large pairing gap limit.
We need further investigation in order to clarify this point.

The resonances which are found at $E=10.60\pm 4.60\hspace{2pt}i$ MeV with $\Delta(r)=0$
move on the trajectories as shown in Fig.\ref{fig6-1} due to the pairing effect.
The trajectory of resonances show that the pairing effect increases both resonance energy and
width, since the imaginary value of the pole on the complex energy plane corresponds to the width
of resonance.

In Fig.\ref{fig7-1}, we show the trajectories of S-matrix all poles existing on the 2nd Riemann
sheet below 20 MeV quasiparticle energy obtained by varying the pairing gap $\bra\Delta\ket$ from
0 to 10 MeV. It seems to be possible to classify the trajectories from their behaviour into
two types except the poles of $d_{5/2}$, $f_{5/2}$ and $f_{7/2}$ denoted by the text in Fig.\ref{fig7-1}.
The $p_{1/2}$ and $p_{3/2}$ poles shown in the upper panel become resonances by the configuration mixing
between hole states existing below the Fermi energy as shown in Fig.\ref{fig0} and the continuum
states due to the pairing effect. The resonance energy and the width are monotonically increased
as the pairing gap is increased. This type of the resonances are called as ``hole-like'' quasiparticle
resonance~\cite{kobayashi}.

The resonances above $E=-\lambda(=8$MeV) are called as the ``particle-like'' quasiparticle resonances
in~\cite{kobayashi}. The resonances at the no pairing limit are formed by the centrifugal barrier and
the spin-orbit potential. Such resonances are called as ``the shape resonance''.
The trajectories in the lower panel above $E=-\lambda(=8$MeV) show the pairing effect on the ``shape
resonance'' except the $d_{5/2}$ resonance which is represented by text in the figure with large width.

The poles of $d_{5/2}$ and $d_{3/2}$  existing below $E=-\lambda(=8$MeV) are anti-bound quasiparticle
states. These poles are moved to the positive direction of the real axis, and become the ``particle-like''
quasiparticle resonance above $E=-\lambda(=8$MeV) at the large pairing limit.
\begin{table}
  \caption{S-matrix poles of hole-like quasiparticle resonances which give the
    clear contribution to the total cross section shown in Fig.\ref{fig10}.}
  \label{table1}
\begin{ruledtabular}
\begin{tabular}{ccc}
$\bra\Delta\ket$ & $p_{1/2}$ & $p_{3/2}$ \\
\colrule
$0.0$ & $11.71$        & $15.68$ \\
$3.0$ & $12.65\pm i  0.28$ & $16.39\pm i 0.08$ \\
\end{tabular}
\end{ruledtabular}
\begin{flushright}
  (For $\lambda=-8.0$ MeV, Unit:MeV)
\end{flushright}
\end{table}
The trajectory's behaviour of the poles of $d_{5/2}$, $f_{5/2}$ and $f_{7/2}$ denoted by the text
in Fig.\ref{fig7-1} is distinct from both ``particle-like'' and ``hole-like'' resonances.
The $d_{5/2}$ resonance may contribute to the nucleon scattering cross section as a background
because it has very wide width. These states may be categorized into ``echo'' state of the nucleon
scattering as suggested by {\it Sasakawa}~\cite{sasakawa}. However, more detailed analysis of
these poles would be necessary to determine the classification.

We show the total neutron elastic scattering cross section as a function of the incident energy
of neutron defined by Eq.(\ref{ei}) in Fig.\ref{fig10}.
The partial cross sections for each angular momentum components are also shown in the same figure.
We use the formulas given by Eqs.(\ref{xsectot}) and (\ref{xsecpart}) to calculate the cross section.
The incident energy $E_i$ is multiplied to the cross section in order to see the contribution of the
partial cross section clearly. The cross sections with $\bra\Delta\ket=0$ and $3.0$ MeV are shown
in the upper and lower panels, respectively. In the lower panel, the total cross section with
$\bra\Delta\ket=0$ MeV is also plotted by the black dashed curve for comparison.

A peak found around $E_i\sim 9.5$MeV in the upper panel is a typical shape resonances formed by
$g_{9/2}$ and $f_{5/2}$ resonances. It is very difficult to see the pairing effect on those resonances
in comparison between the upper and lower panel because those resonances have wide width and
the energy shift is also very small. On the other hand, we can find staggering shapes of
cross section around $E_i\sim 4.5$ and $8$ MeV. Those staggering shapes of cross section are
due to the asymmetric shapes of the partial cross section caused by the hole-like quasiparticle
resonances of $p_{1/2}$ and $p_{3/2}$, respectively. The quasiparticle energies found on
the 2nd Riemann sheet of the complex quasiparticle energy plane are listed in Table \ref{table1}.
The widths for each resonances, $p_{1/2}$ and $p_{3/2}$,  are determined as $\Gamma=0.56, 0.16$ MeV
respectively, because Im $E=\frac{\Gamma}{2}$.

\begin{figure}[htbp]
\includegraphics[scale=0.4,angle=0]{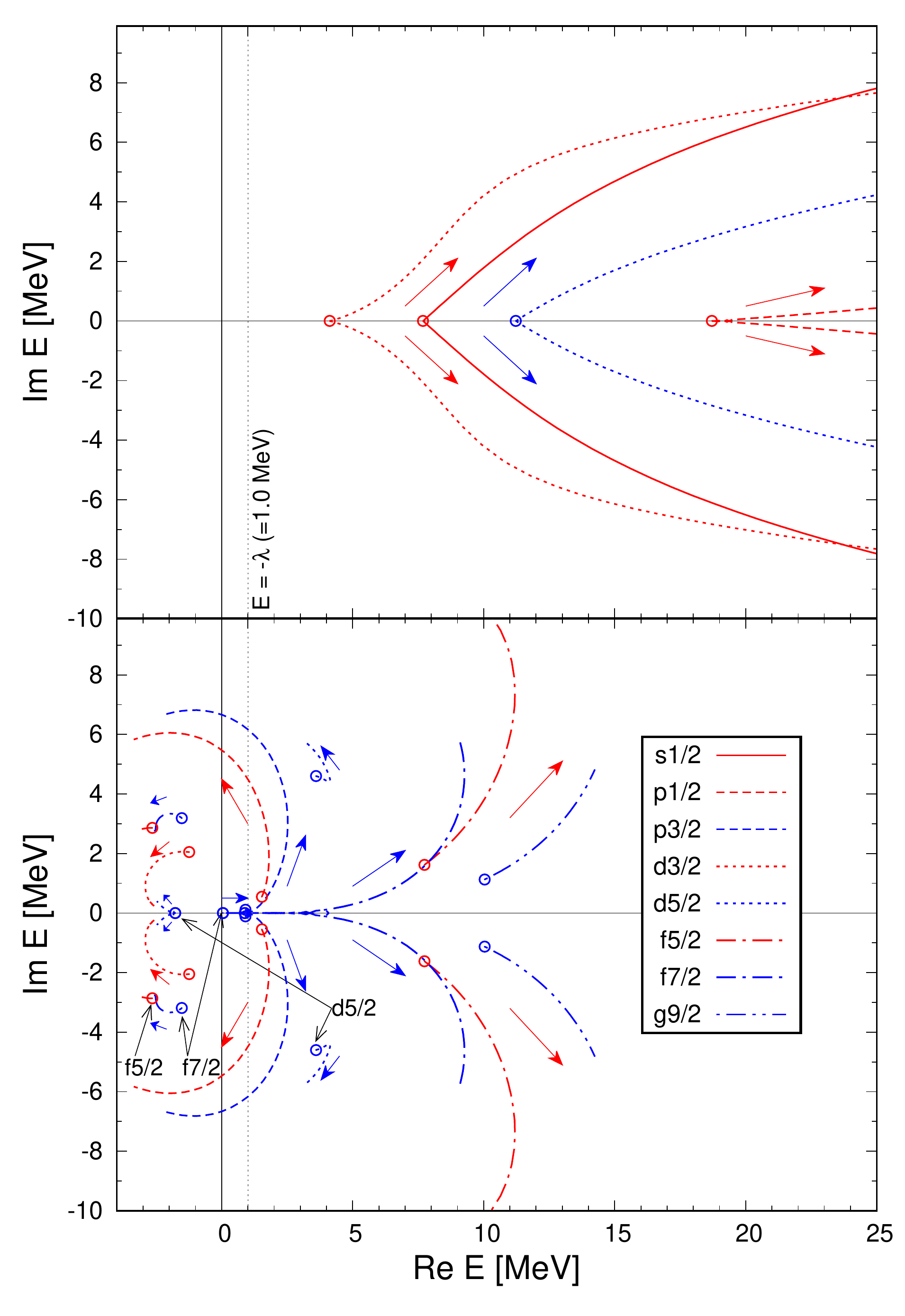}
\caption{(Color online) The same figure with Fig.\ref{fig7-1} but with $\lambda=-1$ MeV.}
\label{fig8-1}
\end{figure}

\begin{table}
  \caption{S-matrix poles of particle-like and hole-like quasiparticle resonances which give the
    clear contribution to the total cross section shown in Fig.\ref{fig11}.}
  \label{table2}
\begin{ruledtabular}
\begin{tabular}{ccc}
\multicolumn{3}{c}{Particle-like}\\
$\bra\Delta\ket$ & $p_{3/2}$ & $f_{7/2}$ \\
\colrule
$0.0$ & $0.89$        & $0.91\pm i 0.11$ \\
$3.0$ & $1.35\pm i  0.17$ & $2.46\pm i 0.01$ \\
\end{tabular}
\begin{tabular}{ccccc}
\multicolumn{5}{c}{Hole-like}\\
$\bra\Delta\ket$ & $d_{3/2}$ & $s_{1/2}$ & $d_{5/2}$ & $p_{1/2}$ \\
\colrule
$0.0$ & $4.12$        & $7.69$       & $11.22$        & $18.70$ \\
$3.0$ & $4.98\pm i  0.26$ & $8.30\pm i 0.48$ & $11.84\pm i  0.33$ & $19.29\pm i 0.03$ \\
\end{tabular}
\end{ruledtabular}
\begin{flushright}
  (For $\lambda=-1.0$ MeV, Unit:MeV)
\end{flushright}
\end{table}

In Fig.\ref{fig8-1}, we show the same plot for the case of $\lambda=-1.0$ MeV to see the trajectories
of the S-matrix poles for the neutron-rich unstable nuclei. The number of the hole-like quasiparticle
resonances shown in the upper panel are increased because of the shallow Fermi energy.
The paring effect on the particle-like quasiparticle resonances ($p_{1/2}$, $p_{3/2}$, $f_{5/2}$ and $g_{9/2}$)
are same qualitatively in shape or form of the trajectories, but more sensitive to the pairing gap
than those shown in Fig.\ref{fig7-1}.
The $d_{5/2}$-resonance indicated by text in the Fig.\ref{fig8-1} is existing with the same width in the no
pairing limit as the one shown in Fig.\ref{fig7-1}. However, the sensitivity for the pairing is quite
different. Almost no pairing effect can be seen in the behaviour of $d_{5/2}$-resonance trajectory
even at the large pairing limit ($\bra\Delta\ket=10.0$ MeV).
\begin{figure}[htbp]
\includegraphics[scale=0.4,angle=0]{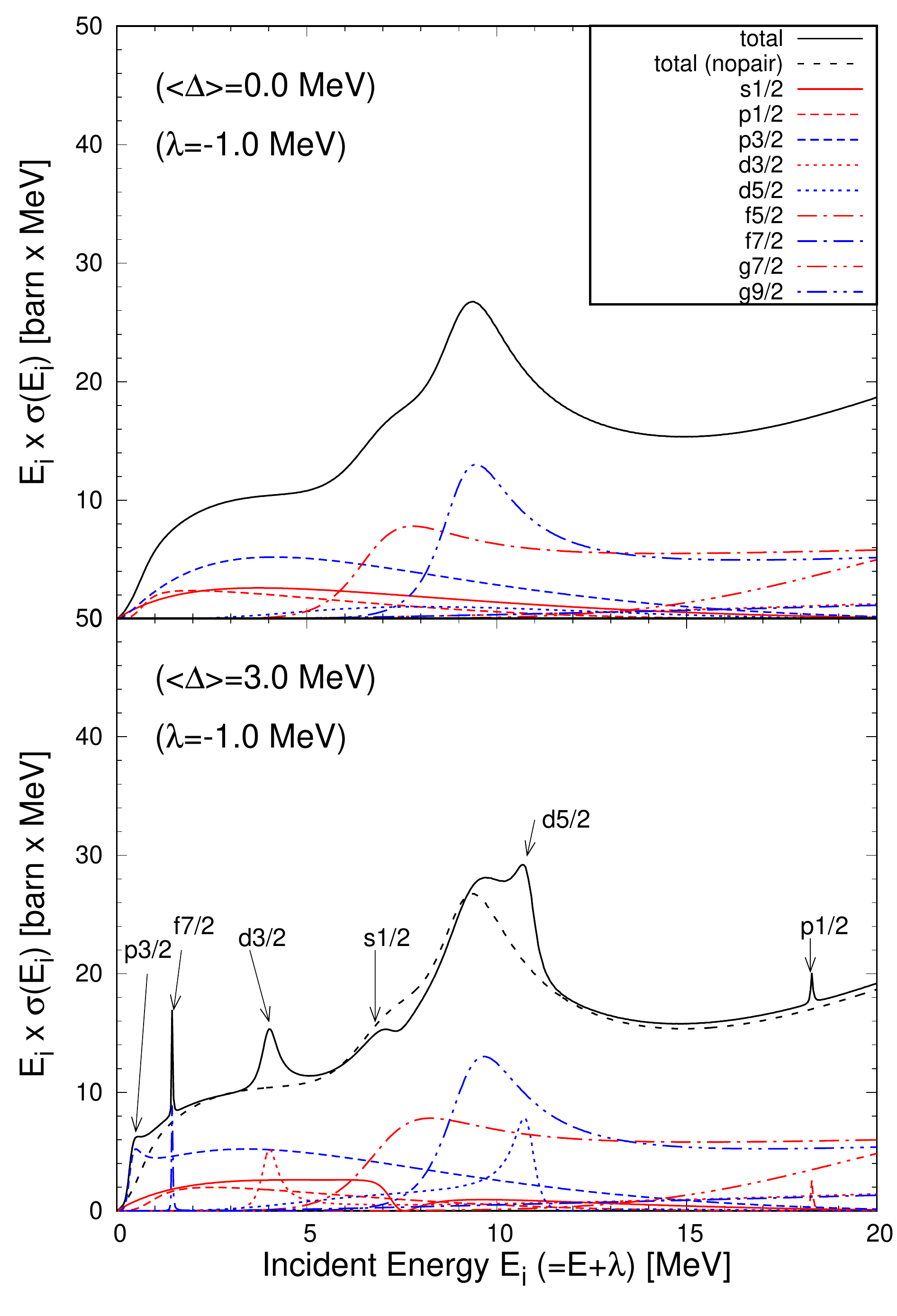}
\caption{(Color online) The total cross section and the partial cross sections for the neutron elastic scattering with $\lambda=-1.0$ MeV.}
\label{fig11}
\end{figure}
The total cross section for the neutron elastic scattering is shown in Fig.\ref{fig11}.
The partial cross sections for each angular momentum are also shown together. In the lower panel,
we can find 6 peaks of resonances which are caused by the pairing effect.
The two peaks indicated as $p_{3/2}$ and $f_{7/2}$ are the particle-like
quasiparticle resonances.
The 4 peaks indicated by text as $s_{1/2}$, $p_{1/2}$, $d_{3/2}$ and $d_{5/2}$ in Fig.\ref{fig8-1} are
the hole-like quasiparticle resonances corresponding the S-matrix poles represented in the upper
panel of Fig.\ref{fig8-1}.
The corresponding quasiparticle energies on the complex quasiparticle energy plane are listed in
Table \ref{table2}. The widths for each resonances can be obtained by multiplying factor $2$ to the
imaginary values of the complex quasiparticle energies.

\section{summary}
We have extended the Jost function formalism based on the Hartree-Fock-Bogoliubov formalism in
order to give the analytic expression of the S-matrix for the nucleon elastic scattering targeting
on the open-shell nucleus. We showed that the S-matrix satisfies the unitarity on the branch-cut
(continuum) in the positive quasiparticle energy region. Adopting the Woods-Saxon potential,
we showed the S-matrix poles on the complex quasiparticle energy/momentum plane by using the
analytic expression of the S-matrix represented by the Jost function.
The trajectories of the S-matrix poles obtained by varying the pairing strength were also
shown for two cases of the Fermi energies given by $\lambda=-8.0$ and $-1.0$ MeV. We pointed out
that the trajectories of resonances can be categorized into two types, the hole-like and
particle-like quasiparticle resonances aside from a couples of exceptions. We confirmed that
such quasiparticle resonances can be observed as the staggering shape or sharp peaks of the total
cross section of the neutron elastic scattering.
We did not compare our results with the experimental data in this study.
For the quantitative discussion, we need to adopt the proper optical potential in order to take
account the proper absorption given by the imaginary part of the optical potential. One of the
proper way is to adopt the microscopic optical potential based on the PVC method. However,
it may be necessary to extend the existing the PVC method to the quasiparticle-vibration coupling
method for the self-consistency of the method although it is a matter of concern that excessive
calculation time is required.
An alternative way is to adopt the GOP, however, readjustment of the parameters
may be necessary because the current parameters have been adjusted without pairing correlation.

\section{Acknowledgments}
We thank Dr.~Yoshihiko Kobayashi (Kyushu University) for useful discussions. This work is funded
by Vietnam National Foundation for Science and Technology Development (NAFOSTED) under grant number
“103.04-2018.303”. T. Dieu Thuy partially thanks the financial support of the Nuclear Physics Research
Group (NP@HU) at Hue University through the Hue University Grant 43/HD-DHH.

\end{document}